\documentclass[amsthm]{llncs}  

\usepackage{amsmath,amssymb}
\usepackage{mathrsfs}
\usepackage{hyperref}

 \usepackage{float}
\usepackage{enumerate}
\usepackage{algorithmic}
 \floatstyle{boxed}
 \newfloat{algo}{h}{loa}
\floatname{algo}{{\textnormal{Algo}}}

\def\l{\ensuremath{\langle}}
\def\r{\ensuremath{\rangle}}

\def\Ldroit{\ensuremath{\mathsf{L}}}
\def\Q{\ensuremath{\mathbb{Q}}}

 \newcommand{\conjug}[1]{\ensuremath{\overline{#1}}}
 \newcommand{\x}[1]{\ensuremath{x_1,\ldots,x_{#1}}}

\newcommand{\wt}[1]{\ensuremath{\widetilde{#1}}}

\newtheorem{Prop}{Proposition}
\newtheorem{Prop*}{Proposition*}
\newtheorem{Theo}{Theorem}

\begin{document}

\begin{frontmatter}
  \title{On the bit-size of non-radical triangular sets} 

 \author{Xavier Dahan\thanks{{\small
       work supported by the JSPS grant Wakate B No. 50567518}}
 }
   
   \institute{Ochanomizu university,
     Faculty of General Educational Research
     \email{dahan.xavier@ocha.ac.jp}
   }


\maketitle

 \begin{abstract}
   We present upper bounds on the bit-size of coefficients
of non-radical lexicographical Gr\"obner bases in purely  triangular form (triangular sets)
of dimension zero. 
This extends a previous work~\cite{DaSc04}, constrained to radical triangular sets;
it follows the same technical steps, based on interpolation.
However, key notion of height of varieties is not available for points with
multiplicities; therefore the bounds obtained are less universal and depend
on input data.
We also introduce a related family of non-monic polynomials that have smaller coefficients,
and smaller bounds.
It is not obvious to compute them from the initial triangular set though.
\end{abstract}


\end{frontmatter}

\section{Introduction}

Triangular sets are the core objects of the triangular decomposition methods to solve polynomial
systems~\cite{LeMMMXi04,Wang2012Elimination,Hu1,chenMMM2012,AuLaMMM99}. Algorithms in this realm are somewhat based on the
generalization of algorithms for {\em univariate} polynomials to multivariate ones, yielding some splittings
viz. ``decomposition''.
The outputs are most of the time {\em regular chains} (a.k.a regular sets).
In dimension zero they can be made particularly simple: they form a reduced lexicographic
Gr\"obner basis $(t_1(x_1),\ t_2(x_1,x_2),\ \ldots ,\ t_n(x_1,\ldots,x_n))$.
We will refer to such a family as a {\em triangular set} in this article.

To solve polynomial systems, it is enough to represent the radical ideal generated by the input polynomials,
thereby most previous works focus on radical triangular sets.
However, triangular sets have the ability to represent some non-radical ideals (called thereafter triangular ideals);
Moreover the radical of the ideal generated by a triangular set is not necessarily triangular, requiring extra work to
decompose it into triangular sets.
If we compare with the Rational Univariate Representation (RUR, see~\cite{Ro99}),
only the multiplicity (which is just a number) of a root is given.
Therefore beyond the theoretical interest,  it is worth studying non-radical triangular sets.

In this article we unveil the {\em structure} and prove upper bounds on the bit-size of coefficients of such triangular sets.
This is an attempt of generalization from radical to non-radical triangular sets, of the results given in~\cite{DaSc04}.
Let us recall briefly the strategy of this paper, since we will follow it.

Step 1) Given the solution points, some interpolation formulas are proved to reconstruct
the triangular set from the points.

Step 2) These formula allows to control the growth of coefficients in function
 of that of the points.

Step 3) A tool from Diophantine geometry called {\em height of variety}
defined through a {\em Chow form} is introduced. It measures somewhat the arithmetic complexity
of the variety, and is endowed of an arithmetic analogue of the B\'ezout theorem (degree of intersection).
This ``Arithmetic B\'ezout theorem'' provides upper bounds in function of {\em any} input polynomial system.

Step 4) A simple modification of the interpolation formulas, called {\em barycentric} form of Lagrange interpolation
defines a family of non-monic polynomials which have smaller coefficients.

We present extensions of Steps 1)-2), and partially 4), to non-radical triangular sets. As for Step 3) the tool (height of variety)
is not available for multiple points. While interpolation in~\cite{DaSc04} is multivariate Lagrange, here it is more general and encompasses
multivariate Hermite interpolation. The input data are not points but primary ideals assumed to be given by a triangular set (see~\eqref{eq:struct}).
The related family of non-monic polynomials mentioned in Step~4) and denoted $N_\ell$ are defined in Theorem~\ref{th:struct}.
In comparison with~\cite{DaSc04}, they seem not easy to compute from the triangular set $T$ (see~\cite{YaDa16})
for an attempt in two variables).
\smallskip

\noindent{\bf Related work.}
For the bit-size bounds, a selection of related work concerned with the RUR, triangular sets, and lexicographic Gr\"obner
bases is~\cite{MaScTs17,DaSc04,Da09,MeSc16}.
The bounds presented are the first ones dealing with non-radical systems having a general type of singularities. Comparatively, a RUR can represent multiplicities (recall that this is just a number) but not a full singularity type.
\smallskip

\noindent{\bf Notation.}
$k$ will denote any field $\Q\subseteq k\subsetneq \conjug{\Q}$. A polynomial ring over $k$, in $n$ variables $x_1,\ldots,x_n$
implicitly ordered such as $x_1\prec x_2\prec\cdots\prec x_n$, a triangular set $T=(T_1(x_1),\ldots,T_n(x_1,\ldots,x_n))$
with $\deg_{x_j}(T_j)=d_j$. Its set of zeros in $\conjug{k}^n$  is denoted $V$.
For a subset $\cal S$ of an arbitrary Cartesian product $E^m$, $\cal S_{\le \ell}$ will denote the projection of $\cal S$ on the first $\ell$ coordinates. 

\section{Interpolation formula}
{\bf Input data.} In the radical case, we want to interpolate points.
Here, the raw input data are primary ideals associated to each solution point in $\conjug{k}^n$.
Thanks to Theorem~2.4 of~\cite{Da17} the primary ideals of a triangular set are
triangular: the lexicographic Gr\"obner basis of these primary ideals are triangular sets.
Still over $\conjug{k}$, a prime ideal associated with $\l T \r$ is of the form $\l x_1-\alpha_1,
\ldots, x_n -\alpha_n\r$, for a solution point $(\alpha_1,\ldots,\alpha_n)$.
The corresponding primary ideal $\l t^{(\alpha)}\r$ has the following shape (Proposition~2.2 of \cite{Da17}):
First 
$t_1^{(\alpha)}(x_1)  = (x_1-\alpha_1)^{\delta_1(\alpha)}$ and  
in general $t_{n}^{(\alpha)}(x_1,\ldots,x_n)$ is equal to:
\begin{equation} \label{eq:struct}
(x_n-\alpha_n) ^{\delta_n(\alpha)} +  \sum_{i_1=0}^{\delta_1(\alpha_1)-1} 
\sum_{i_{n-1}=0}^{\delta_{n-1}(\alpha)-1}
\sum_{i_n=0}^{\delta_n(\alpha)-1}   c_\alpha[i_1,\ldots,i_n]  \prod_{j=1}^{n}  (x_j-\alpha_j) ^{i_j} 
\end{equation}
where: 
(i) For $1\le u\le n$,  $\deg_{x_u}(c[i_1,\ldots,i_n])< \delta_u(\alpha)$,

\noindent (ii)  $c[0,\ldots,0,i_\ell]=0$ for all $i_\ell < \delta_\ell(\alpha)$ and for $\ell=2,\ldots,n$.

\noindent (iii) Note that Taylor expansion gives:
$
c[i_1,\ldots,i_n]= \frac 1 {i_1 ! \cdots i_{n}!} \frac{\partial^{i_1+\cdots+i_{n}}t_{n}}
{\partial x_1 ^{i_1} \cdots \partial x_{n} ^{i_{n}}} (\alpha_1,\ldots,\alpha_{n})
$.
 

We denote by $T_{\ell+1}[\alpha]$ the polynomial $T_{\ell+1}  \bmod \l t_{\le \ell}^{(\alpha)} \r $.
Theorem~3.1 of~\cite{Da17}, the ring $\bar{k}[\x{n}]/ \l t^{(\alpha)}_{\le \ell}\r$ is {\em Henselian}.
Hence $T_{\ell+1}[\alpha]$ admits a unique factorization as follows:
\begin{equation}\label{eq:UF}
T_{\ell+1}[\alpha]\equiv \prod_{\beta\in V_{\le \ell+1}}  t_{\ell+1}^{(\beta)} \ \bmod \l t^{(\alpha)}_{\le \ell}\r ,
\ \  \text{where} \ \ 
(\beta_1,\ldots,\beta_\ell)=(\alpha_1,\ldots,\alpha_\ell).
\end{equation}
and
$t_{\ell+1}^{(\beta)}=(x_{\ell+1} -\beta_{\ell+1})^{\delta_{\ell+1}(\beta)} + \sum_{i_1,\cdots,i_{\ell},r}  c_\beta[i_1,\ldots,i_{\ell},r]
(x_{\ell+1} -\beta_{\ell+1})^r \cdot \prod_{j=1}^{\ell} (x_j - \alpha_j)^{i_j}$ for some $c_\beta[i_1,\ldots,i_{\ell},r]\in\bar{k}$.
This key result allows to prove Proposition \ref{prop:idem}.

\noindent{\bf Notation.} A sequence $\alpha^1\in V_{\le 1} \ , \ \ \alpha^2\in V_{\le 2}\ ,\ \ldots\ ,\ \ \alpha^\ell\in V_{\le \ell}$
will {\em not} denote ``$j$-th power of $\alpha$'',
but  points $\alpha^j=(\alpha_1^j,\ldots,\alpha_j^j)$ with the additional convention that
$(\alpha^j_1,\ldots,\alpha^j_j)=(\alpha^{j+1}_1,\ldots,\alpha^{j+1}_j)$. 
We say that $\alpha^{j+1}$ {\em extends} $\alpha^j$.
\begin{Prop}\label{prop:idem}
Let $\gamma \in V_{\le \ell+1}$ be a root that extends
$\alpha = (\alpha_1,\ldots,\alpha_{\ell})\in V_{\le \ell}$.
\begin{enumerate}
\item\label{idem1}  
$e_{\ell+1}(\gamma)
\equiv \frac{T_{\ell+1}[\alpha]}{t^{(\gamma)}_{ \ell+1}}
\bmod \l t^{(\alpha)}_{\le \ell}\r $
is a polynomial
in $(\conjug{k}[\x{\ell}]/\l t^{(\alpha)}_{\le \ell} \r)[x_{\ell+1}]$
\item\label{idem2} Orthogonality: Given $\beta\not=\gamma\in V_{\le \ell+1}$:\\
\centerline{$
e_{\ell+1}(\beta) \cdot e_{\ell+1}(\gamma) = 0 \text{~in~}
A_\alpha:=\conjug{k}[\x{\ell+1}]/\l t_1^{(\alpha)},\ldots, t^{(\alpha)}_{\ell},
T_{\ell+1}[\alpha]\r.
$ }
\item\label{idem2bis} $e_{\ell+1}(\beta) \equiv 0 \bmod \l t_{\le \ell+1}^{(\beta')} \r$ if $\beta'\not=\beta$,
\hfill  and $\quad \wt{e_{\ell+1}}(\gamma)\equiv 1 \bmod \l t_{\le \ell+1}^{(\gamma)}\r$.
\item\label{idem3} There are polynomials $u_{\ell+1}(\gamma)$ and $v$
such that \\
$ u_{\ell+1}(\gamma) e_{\ell+1}(\gamma) + v \, t_{\ell+1}^{(\gamma)} \equiv 1 \bmod \l t_{\le \ell}^{(\alpha)}\r$, with  $\deg_{x_{\ell+1}}( u_{\ell+1}(\gamma) ) < \deg_{x_{\ell+1}}(t_{\ell+1}^{(\gamma)})$.

Denote $\widetilde{e_{\ell+1}} (\gamma)\equiv u_{\ell+1}(\gamma) e_{\ell+1}(\gamma) \bmod
 \l t_{\le \ell}^{(\alpha)}\r $.
The family $\{ \widetilde{e_{\ell+1}} (\gamma)\}_{ \gamma}$ is a
complete family of orthogonal idempotents of the algebra $A_\alpha$.
\end{enumerate}
\end{Prop}

Have in mind that Lagrange interpolation use idempotents.
\begin{Theo}\label{th:struct}
Write $t^{(\alpha)}_{\le \ell}=(t_1^{(\alpha)},\ldots,t_\ell^{(\alpha)})$ the triangular sets defining the primary ideal of
associated prime  $\l x_1-\alpha_1,\ldots,x_{\ell+1} -\alpha_{\ell+1} \r$.
\begin{equation}\label{eq:T}
T_{\ell+1} \equiv \sum_{\alpha^1\in V_{\le 1}}\sum_{\alpha^2\in V_{\le 2}}
\cdots \!\!\!\!\!\! \sum_{\alpha^\ell\in V_{\le \ell} } \!\!\!
\widetilde{e_1}(\alpha^1)
\cdots \widetilde{e_{\ell}}(\alpha^\ell) 
\cdot T_{\ell+1}[\alpha^\ell] \bmod \l t^{(\alpha)}_{\le \ell}\r.
\end{equation}
where it is assumed that $\alpha^{j+1} \in V_{\le j+1}$ extends $\alpha^{j}\in V_{\le j}$.
We define $N_{\ell+1} $ by the {\em same} formula, using polynomials $e_i(\alpha^i)$ instead of $\wt{e_i}(\alpha^i)$. 
\end{Theo}
Since the family of idempotents used to define $T_{\ell+1}$ is complete, $T_{\ell+1}$ is monic. This is not the case for $N_{\ell+1}$.
Therefore they cannot be used to perform {\em reduction} through a division,
but their interest lies in their small coefficients, and in that they generate the same ideal
as $\l T_1,\ldots,T_{\ell}\r$ hence encodes the same information; In the radical case
they were used in conjunction of modular methods. 

A natural  question is whether
we can compute the polynomials $N_\ell$'s from the $T_\ell$'s.
The answer is not trivial and not addressed in these pages. However it is not unreasonable
to expect an almost linear complexity algorithm to compute it, as shown for the case of
two variables in~\cite{YaDa16}. It boils down to compute the polynomial denoted  $F_{\ell+1}$ hereunder
\begin{Prop}\label{prop:F}
We have  $F_{\ell+1} T_{\ell+1} \equiv N_{\ell+1} \bmod \l T_1,\ldots,T_{\ell}\r$, with:\\
\centerline{$F_{\ell+1}:= 
\sum_{\alpha^1\in V_{\le 1}}\sum_{\alpha^2\in V_{\le 2}}
\cdots \sum_{\alpha^\ell\in V_{\le \ell}} 
\!\!  e_1(\alpha^1)
\cdots e_{\ell}(\alpha^\ell)$}\\
and where  the same convention on $\alpha^1,\alpha^2,\ldots$ as in Theorem~\ref{th:struct} is adopted.
\end{Prop}
In the radical case, it is easy to show that $F_{\ell}=\frac{\partial T_1}{\partial x_1}
\cdots  \frac{\partial T_\ell}{\partial x_\ell}$.

\section{Bit-size consideration}
\noindent {\bf Preliminary.}
This last section states and comments on some upper-bounds on the bit-size of coefficients in  $\Q$ appearing
in the polynomials $T$ and $N$. The difficulties compared to the radical case
are first, that the interpolation formulas are more complicated to handle,
and second, that there is no notion of Chow form, yet of notion height of varieties,
in our non-radical context;
Whereas it was a key tool in~\cite{DaSc04} to obtain {\em intrinsic} bit-size bounds
(see Step 3) in Introduction).
The upper bounds that we can obtain are less universal: we assume that
the input primary ideals are given in triangular form, which we have written
$t^{(\alpha)}$, while universal bounds would not need this assumption
(another input for the primary ideals may well have smaller coefficients,
hence yield better bounds). 
However bounds~\eqref{eq:NT}  give a reasonable indication on the bit-size,
and are also of interest to understand the growth of coefficients
in multivariate interpolation with singularities, including Hermite's as a subcase.
\smallskip

\noindent {\bf Statement.}
We  use the formalism of {\em height} of polynomials
classical in Diophantine approximation theory.
The notation $h(f)$ denotes the height
of the polynomial $f$, and can be thought as the max bit-size of its coefficients.
Recall that the input ``raw'' data are the triangular set $\{t^{(\alpha)}, \ \alpha\in V\}$
generating the primary ideals of $\l T\r$. With the notations of~\eqref{eq:struct}, this includes 
the exponents $\delta_i(\alpha)$ and coefficients $c_\alpha[i_1,\ldots,i_\ell]$.
We define:
\begin{equation}\label{eq:H}
H_\ell(\beta^\ell):= \max_{i_1,\ldots,i_\ell} h(c_\beta[i_1,\ldots,i_\ell]) 
+ i_1 h(\beta_1) + \cdots + i_\ell h(\beta_\ell),
\end{equation}
and $\Ldroit_\ell(T):= \max_{\alpha \in V_{\le \ell} } (\sum_{i=1}^\ell H_i(\beta^i))$.
Denote by $\mu_\ell(\beta^\ell):= \delta_1(\beta_1) \cdots \delta_\ell(\beta_\ell)$
the local multiplicity at $\beta^\ell$, and finally:
\begin{equation}\label{eq:bounds}
H_\ell (T) :=\sum_{\beta^\ell \in V_{\le \ell}} H_\ell (\beta^\ell),
\quad
 \mu_\ell(T) := \max_{\beta^\ell \in V_{\le \ell}}  \mu_\ell(\beta^\ell),
\quad D_\ell := \sum_{i\le \ell} d_i
\end{equation}

The bit-sizes of any coefficient of $N_{\ell+1}$, or $T_{\ell+1}$, are lower than  quantities  whose
dominating terms are respectively:
\begin{equation}\label{eq:NT}
  H_{\ell+1}(T) + \widetilde{O}( \Ldroit_\ell (T) \cdot D_{\ell+1} \cdot \mu_\ell(T)  ), \qquad
D_{\ell+1} H_{\ell+1}(T) + \widetilde{O}( \Ldroit_\ell (T) \cdot D_{\ell+1}^2  \cdot  \mu_\ell(T)   ) 
\end{equation}

\noindent {\bf Rationale.} $\widetilde{O}(\, . \, )$ is a big-Oh notation that hides any additional logarithmic factors.
The quantity $H_\ell(\beta^\ell)$ of ~\eqref{eq:H} should be thought as a generalization 
to primary components of the height of a projective point:
coefficients and exponents in the Taylor expansion~\eqref{eq:struct} are simply
``naturally'' taken into  account. If the point is simple,
then this quantity coincides with the ``traditional'' height of a point. In addition,
the quantity $H_\ell(T)$ of~\eqref{eq:bounds} reflects a certain additivity of
the ``height''
under distinct primary components; this also generalizes the additivity of the height of varieties under disjoint union. 
Therefore, the quantities involved are ``natural'' extension to those more traditional
used in the case of simple points. The occurrence of quantities like $\Ldroit_\ell(T)$ or
the multiplicity at a point $\mu_\ell(\beta^\ell)$ is due to 
technical details due to the presence of multiplicities.

If $T$ generates a {\em radical} ideal, then
$\mu_\ell(\beta^\ell)=1$, $H_\ell(\beta^\ell)=h(\beta^\ell)$; 
moreover $H_\ell(T)\approx h(V_{\le \ell})$, $D_\ell \le \deg(V_{\le \ell})$ which are respectively
the {\em height} of the variety $V_{\le \ell}$ and its degree, and
if we discard the
values $\Ldroit_\ell(T)$  then the bounds in~\eqref{eq:NT}
become  roughly $h(V_\ell) +  \widetilde{O}(\deg(V)  )$ for $N_{\ell+1}$,
and  $\deg(V_\ell) h(V_\ell) +    \widetilde{O}(  \deg(V_{\le \ell})^2)$
for $T_{\ell+1}$.
These bounds are similar to the ones obtained in~\cite{DaSc04}.
This shows that the bounds~\eqref{eq:NT}
``faithfully''  extend the ones of the radical case.
The difference between the bounds for $N_{\ell+1}$ and for $T_{\ell+1}$ is roughly the factor $D_\ell$, and this ratio is comparable to that of~\cite{DaSc04}.

Finally, experiments not reported here (but see  Table~1 in~\cite{YaDa16} for some data
in two variables) show that the size of polynomials $T_\ell$
can be dramatically larger than $N_\ell$'s.


\bibliographystyle{plain}

\end{document}